# Quantum Critical Point, Scaling and Universality in High $T_c$ $(Ca_xLa_{1-x})(Ba_{2-c-x}La_{c+x})Cu_3O_y$

B. Watkins[2], H. Chashka[1], Y. Direktovich[1], A. Knizhnik[1], and Y. Eckstein[1,2]

[1] *Department of Physics, Technion, Haifa Israel*

[2] *Department of Physics, Northwestern University, Evanston, IL 60208 USA*

Using charge transport observations on sintered ceramic samples of CLBLCO we failed to observe the Quantum Critical Point (QCP) where it is expected. Experimental data relating Cooper pair density, electrical conductivity, and superconductivity critical temperature suggest that Homes' relation might needs a more specific definition of 'σ'. Transport observations on YBCO single crystals will resolve this question.

At present, we still do not understand many features of the phenomenon of high temperature superconductivity (HiT$_c$). We do know that Cooper pairs are formed, and there is a good deal of experimental evidence that the pairing interaction has d-type symmetry. But it is also clear that in the underdoped region of the phase diagram the resistivity vs. temperature behavior is anomalous and it is generally assumed that in the extremely overdoped region, the HiT$_c$ materials become Fermi Liquids. However, at present, there is no clear-cut experimental evidence to support this assumption.

In view of this lack of knowledge, it seems worthwhile to search the properties of all HiT$_c$ compounds for universal general features. For example, it is known that all compounds have a "pseudogap" region above T$_c$. More quantitatively, Uemera [1] has shown that in the underdoped region all the HiT$_c$ compounds have the same $n_s/m^*$ for the same T$_c$ (where $n_s$ is the density of Cooper pairs and $m^*$ is their effective mass). More recently it has been suggested [2] that a new universal scaling relation exists: $n_s = A\sigma T_c$ (where A is a constant, σ is the dc conductivity, and T$_c$ is the superconductivity transition temperature). Another suggested general feature is that a quantum critical point (QCP) exists, and that it is located slightly above the optimum T$_c$. In this paper we experimentally investigate the latter two suggestions: existence and location of QCP and $n_s = A\sigma T_c$. It is shown that neither prediction is valid for CLBLCO. The applicability of the latter relation might depend upon the specific definition of σ. We speculate that small differences in structure cause the difference between CLBLCO and the other compounds particularly YBCO and (CA,Y)BCO.

Here CLBLCO denotes a family of 1-2-3 compounds of composition $(Ca_xLa_{1-x})(Ba_{2-c-x}La_{c+x})Cu_3O_y$. In these materials, $Ca_xLa_{1-x}$ replaces the yttrium and $Ba_{2-c-x}La_{c+x}$ replaces the $Ba_2$ of the parent compound YBCO ($YBa_2Cu_3O_y$).

The present paper is primarily concerned with transport properties of the CLBLCO family of compounds and with any implications that these transport properties may have upon the existence of a QCP. This family of compounds differs markedly from the parent compound YBCO. One important difference is that they remain tetragonal over the entire range of oxygen doping (in contrast to YBCO, which is orthorhombic in the superconducting region). Another, very advantageous feature is that, in contrast to YBCO, it is possible to cover the entire range from strongly underdoped to well past the optimum by varying only the oxygen content [3]. Let us define Q as the total oxidation state of all the cations except copper: Q=7+c. We note that Q does not depend on x. From the stoichiometry, we see that the average oxidation state of the copper is (2y-7-c)/3, but we do not know how many holes reside on the $CuO_2$ planes.

Most of the previous work of our group was for Q=7.25. [i.e. for $(Ca_xLa_{1-x})(Ba_{1.75-x}La_{0.25+x})Cu_3O_y$]. In the present paper, we will also discuss the case Q=7.5 [i.e. $(Ca_xLa_{1-x})(Ba_{1.5-x}La_{0.5+x})Cu_3O_y$]. Fig. 1 shows $T_c$ as a function of the oxygen doping, y, for the case Q=7.25. The particular case x=0.1 has the useful property that it is not too difficult to vary the doping level all the way from the underdoped non SC region through a SC region to another non SC overdoped phase, by changing only the oxygen concentration, y. It should be emphasized that there is no crystallographic phase transition over the entire doping region, from normal underdoped to SC to normal overdoped [3].

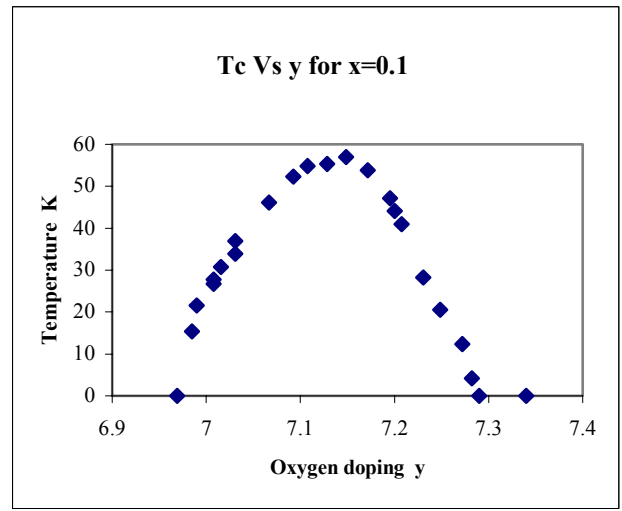

**Fig. 1.** $(Ca_xLa_{1-x})(Ba_{1.75-x}La_{0.25+x})Cu_3O_y$. T$_c$ as function of oxygen doping.

These characteristics give us a unique opportunity to study and compare properties of a family of $HiT_c$ compounds in the underdoped and overdoped regions. In particular it is possible to see whether the extreme overdoped phase is a Fermi liquid, and also to see whether there is a QCP in the SC "dome". Moreover, it is possible to undertake these investigations without requiring a huge magnetic field to suppress superconductivity. Sachdev [4] has suggested that in these complicated compounds there are competing states very close to the ground state, so that changing a parameter can cause dramatic qualitative changes. For example [5] there exists at zero temperature a critical doping level, $y_c$, below which the material is an insulator and above which the material is a conductor. This $y_c$ defines the quantum critical point (QCP). At finite temperature one might expect that below some temperature the electrical resistivity becomes independent of temperature for the highly doped compound ($y>y_c$) and that resistivity increases below T*, (i.e. dR/dT<0 for T<T*) for low doping ($y<y_c$). Recently, Dagan et al. [6] have observed such behavior in the n-type material PCCO and have found that there is a QCP in the SC region (they used a high magnetic field to suppress SC). It is well known that the very underdoped p-type $HiT_c$ superconductors exhibit rises in resistivity (dR/dT<0) as temperature approaches the sample superconductivity transition, $T_c$. Boebinger et. al [7] using a very high magnetic field (61 T) on $La_{2-x}Sr_xCuO_4$ found that there is an $x_0$ such that for $x<x_0$ the compound will be an insulator at zero temperature and for $x>x_0$ it will be metallic. (One should note that $x_0$ is different for the ab directions than for the c direction.) Tallon et. al. [8] used other experimental data as evidence for the existence of a QCP close to the same doping value, $x_0$, and they suggested this to be a universal property of $HiT_c$ compounds.

We have fabricated ceramic samples and have observed their resistivities. The method of preparation has been described in previous publications [9]. The important feature is that by annealing the samples in an oxygen atmosphere at various temperatures and various oxygen pressures one can vary the oxygen concentration, y, defined to be the doping level. An equally important fact is that one can measure the oxygen content y by an independent (chemical) procedure [10]. This gives an accurate measure of the total doping (but of course, it does not give the number of mobile holes in the $CuO_2$ plane).

Typical sample dimensions were 10 mm long and 2 mm wide. The thicknesses were reduced by abrasive grinding from the press dimension of 2 mm to 0.2 or 0.3 mm needed for accurate measurements of the Hall coefficients. (Hall effect data are not presented in this work.) Six contacts (two current, two resistivity, and two Hall effect) were made to the samples by pressing indium into the respective surfaces and indium-soldering to the resulting contacts. 100 mA current pulses of duration 250 - 300 ms were injected along the sample axis while V=IR (and transverse Hall effect) potential differences were recorded. To isolate and compensate thermally generated voltmeter lead potentials, -100 mA current pulses were also injected and observed. The time interval between successive pulses was from 2 to 20 seconds (proportional to the observed heating). Every datum represents at least 10 pulses of each current polarity that were observed and treated statistically. Periodically, pulse amplitudes as small as 10 mA were applied making us confident that no amplitude (or heating) anomalies are observed. Data were recorded from samples having various oxygen doping levels.

In the overdoped region the resistivity was found to be linear with temperature similar to the behavior of other $HiT_c$ compounds. In the underdoped region dR/dT becomes negative below some temperature, T*, (but still above $T_c$, i.e. before entering the SC region). The data in the extreme overdoped region were surprising. Fig. 2 shows resistivity data for y=7.338 and x= 0.1, for the material $(Ca_xLa_{1-x})(Ba_{1.75-x}La_{0.25+x})Cu_3O_y$. In all our extremely overdoped samples dR/dT was found to be negative below 30K. We suspected that we might possibly have been observing a Kondo effect. We therefore made new samples using the purest ingredients commercially available: all of 99.9999% purity where available or 99.999% purity with iron or nickel 1-2 parts in $10^6$ otherwise. The data in Fig. 2 was gathered from a sample made from these high-purity components. From 30K (the temperature of minimum resistivity) to 90K, the resistance R varies as $T^2$ as expected for a Fermi liquid. However, we believe that this is not true Fermi liquid behavior, because (a) the temperature is well above the range where Fermi liquid behavior is expected, and (b) the resistivity goes through a *minimum* at lower temperatures. We believe that the $T^2$ fit reflects the parabolic first approximation of every smooth minimum. These data also raise doubts about the existence of QCP in the superconducting region, but leave open the possibility that a QCP may exist at higher doping levels.

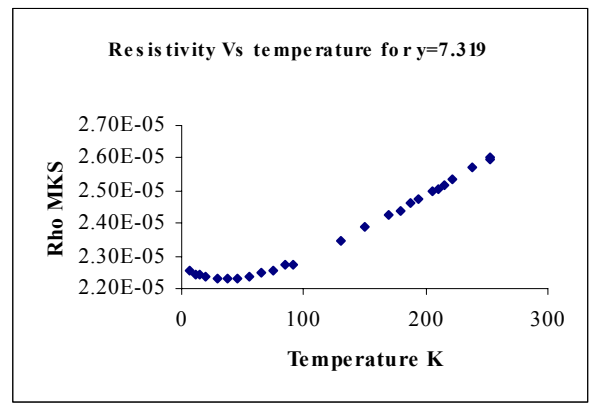

**Fig. 2.** $(Ca_xLa_{1-x})(Ba_{1.75-x}La_{0.25+x})Cu_3O_y$ Resistivity as function of temperature for y= 7.331.

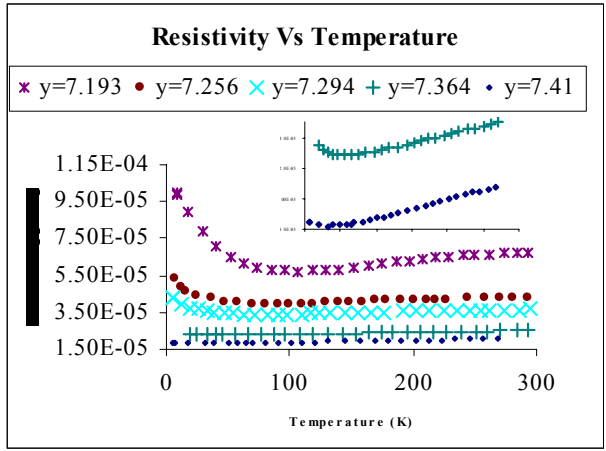

**Fig. 3.** $(Ca_xLa_{1-x})(Ba_{1.5-x}La_{0.5+x})Cu_3O_y$. Resistivity as function of temperature for different y (doping) values. Inset: High doping part of the curves in Fig. 3 emphasizing the occurrence of minima in R(T).

The question concerning what happens in the superconducting region when the SC is suppressed remains open. We were unable to investigate this, since we have no access to sufficiently high magnetic fields. Moreover, the effects of huge magnetic fields have not yet been studied for this compound.

Members of the CLBLCO family with Q=7.5 [i.e. materials with composition

$(Ca_xLa_{1-x})(Ba_{1.5-x}La_{0.5+x})Cu_3O_y$] were also investigated [11] [12]. It was found that except for the case x=0.5 they have the same structure as the Q=7.25 materials discussed above. The calcium always occupies yttrium sites, but the lanthanum is located at two places; it replaces the yttrium at some sites and it replaces the barium at other sites. The structure remains tetragonal for all observed compounds (all x and all y). For x>0.1 there is a SC region. Unfortunately the published data do not reveal the full curve as function of both x and y. We have published data [13] for x=0.4 and Q=7.25, 7.35 and 7.45 for various y. We note that for Q=7.25 the optimum doping level, $y_{opt}$, does not depend upon x and we can assume that this also applies to other Q values. A slight extrapolation of $y_{opt}$ versus Q suggests that for Q=7.5 $y_{opt}$ ≈7.27. We know that the range of doping levels in the superconducting "dome" for Q=7.25 and x=0.1 is 6.97≤y≤7.29. It seems reasonable that the range of the dome might be similar for Q=7.5 and x=0.1 i.e. 7.11≤y≤7.42. However, for Q=7.5 and x=0, the material has no SC region [i.e. $La(Ba_{1.5}La_{0.5})Cu_3O_y$ is never SC for any y]. It is very probable that its doping range covers the entire range of the SC dome for the case x=0.1 and Q=7.5. Fig. 3 shows the temperature dependence of the resistivity for samples with various values of y. These curves each show that there is a temperature, T*, at which the resistance R has a minimum; below T* dR/dT is negative. Furthermore, dT*/dy is also negative, i.e. T* falls as y increases. These facts seem to answer the question about the existence of a QCP inside the SC dome. At least in the case where the SC dome is absent, there is no sign of QCP. Either other effects mask evidence of any QCP or QCP does not appear in CLBLCO at the superconducting range.. It is not clear whether this is due to the fact that the location of the QCP is not universal or is due to the unique properties of CLBLCO. One may ask whether single crystals of CLBLCO would exhibit behavior different from polycrystalline samples with respect to a QCP. This problem was treated theoretically for the general case by Stroud [17] and experimentally for the HiT$_c$ material $La_{2-x}Sr_xCuO_4$ [18]. It was shown that the temperature dependence of the resistivity is the same for sintered pellets as for single crystals except for the fact that ρ is larger by a factor of two to three in the polycrystalline samples. Our lack of single crystals *should* not affect our search for a QCP.

CLBLCO is unique in several respects:
**(1)** Underdoped and overdoped specimens with the same $T_c$ have the same buckling angle [14]. The buckling angle is the angle between copper and the two adjacent oxygen atoms in the $CuO_2$ plane.
**(2)** Underdoped and overdoped samples [15] show the same pressure dependence of $T_c$.
**(3)** Underdoped and overdoped specimens with the same $T_c$ have the same density, $n_s/m^*$, of SC pairs and obey Umera's rule [16]
**(4)** In the underdoped and overdoped regions the resistance R(T) always has a minimum as function of temperature**.**

Umera's rule above (3) means that CLBLCO cannot obey Homes' proposed universality relation [2] for two reasons:

(a) For the same $T_c$, μSR shows $n_s$ of underdoped samples equals $n_s$ of overdoped samples and
(b) The conductivity is larger in the overdoped samples than in the underdoped samples.

We note that unlike the YBCO case, the "chain" in CLBLCO oxide is actually a plane (the sample is tetragonal) and 2D bond percolation theory [19] predicts that once the number of oxygen atoms is larger than half the number of coppers (y>7) this "chain" plane should begin to conduct electricity.

There are three possibilities:
**(1)** The addition of oxygen might increase the charge in the "chain" plane and decrease the charge in the $CuO_2$ plane. This implies that the sample effectively becomes "underdoped" when $y>y_{opt}$. This response is consistent with the μSR observation that $n_s$ is lower than optimum in samples with $y>y_{opt}$. One may note that conductivity in the "chain" plane can allow $n_s=A\sigma_{CuO(2)}T_c$ where the conductivity ($\sigma=\sigma_{CuO(2)}+\sigma_{Chain}$) now has $CuO_2$ plane and "chain" plane components. However, bond valence sum (BVS) calculations [20] indicate that the number of holes in both types of planes increases monotonically with oxygen doping; this scenario therefore seems improbable. NMR experiment on this compound will be able to measure directly the number of holes in the $CuO_2$ plane.
**(2)** Maybe, for some reason, the holes in the "chain" plane are not mobile and therefore the conductance is only due to the $CuO_2$ planes. In this case Homes' relation cannot hold for the present compound. This possibility would make us question our understanding of YBCO chain conduction.
**(3)** Most probably, there is an increase of mobile holes in both "chain" and $CuO_2$ types of Cu planes. This model is consistent with BVS calculations. In this case $n_s=A\sigma_{CuO(2)}T_c$ is definitely possible.

Assuming that possibility (3) occurs, Homes' rule tells us the contribution to the electrical conductivity of the $CuO_2$ in the underdoped region and from the equality of $n_s$ we may infer σ of the $CuO_2$ plane in the overdoped region as well. With the measurement of the total σ we effectively measure the conductivity of the "chain" plane. This model can be checked in single crystals of YBCO. In YBCO there is a range of oxygen doping where the chain contributes and the current is parallel to the <u>b</u> direction. Differences in conductivity between the a and b directions in YBCO isolate the chain conductivity from the $CuO_2$ plane conductivity.

CLBLCO is a 1-2-3 compound in which some of the atoms in YBCO are replaced by other elements. When some of the yttrium is replaced by calcium, the changes are not drastic. These alloys can then be doped to some extent into the overdoped region, but the crystal structure remains unchanged and orthorhombic. In CLBLCO, the lanthanum replaces some of the barium *and* some of the yttrium. As was shown by neutron diffraction [14] and by BVS calculation [20], these substitutions also involve some structural changes. The main change is that the apical oxygen $O_4$ is not always at the exact position it would have occupied in YBCO. $O_4$'s position now depends on its neighbors. If both neighbors are Ba or both are La, the apical oxygen's position is unchanged from YBCO; but if one neighbor is Ba and the other La, the oxygen is displaced towards the smaller La atom. This is presumably connected with the fact that this compound is always tetragonal and can be oxygen-doped over the entire range of SC materials. It is known that the polarizability of the apical oxygen is very high and that this oxygen can easily change its state [21] (from $sp_z$ to $sp^3$). It might be possible that such a change in orbital state and structural deformation stress might contribute to charge localization and might mask the existence of a quantum critical point. This hypothesis might be checked by observing the resistivity at lower (dilution refrigerator) temperatures; however, if the resistance becomes too high, it may become difficult to disentangle the mixture of c-axis and ab plane contributions. Therefore, despite its experimental difficulty, it would be very desirable to grow single crystals of CLBLCO for future study.

In conclusion it is clear that Homes' universal relation $n_s=A\sigma T_c$ needs to be tested for the case that the conduction is not limited to the $CuO_2$ plane. Additionally, the QCP may not always be located close to the optimum doping level, and might even not exist in some HiT$_c$ materials. A striking new feature of CLBLCO is the appearance of unexpected similarities between the underdoped and overdoped regions. For the same $T_c$ we find that:

(1) The buckling angles are equal,
(2) The pressure dependences of $T_c$ are the same,
(3) The ratio $n_s/m^*$ are equal, and for the normal part:
(4) For all doping levels, there exists a T* such that for T<T*, dR/dT<0.

All of these properties raise the possibility that CLBLCO does not belong to the same class of materials as the cuprates discussed by Homes [2], by Dagan [6], by Boebinger [7], and by Tallon [8].